# On the attenuation coefficient of monomode periodic waveguides

A. Baron[1], S. Mazoyer[1], W. Smigaj[1] and P. Lalanne[1,2]

[1] *Laboratoire Charles Fabry de l'Institut d'Optique, CNRS, Univ Paris-Sud, Campus Polytechnique, 91127 Palaiseau cedex, France*

[2] *Laboratoire Photonique, Numérique et Nanosciences, Université Bordeaux 1, CNRS, Institut d'Optique, 33405 Talence cedex, France.*

It is widely accepted that, on ensemble average, the transmission $T$ of guided modes decays exponentially with the waveguide length $L$ due to small imperfections, leading to the important figure of merit defined as the attenuation–rate coefficient $\alpha = -\langle\ln(T)\rangle/L$. In this letter, we evidence that the exponential–damping law is not valid in general for periodic monomode waveguides, especially as the group velocity decreases. This result that contradicts common beliefs and experimental practices aiming at measuring $\alpha$ is supported by a theoretical study of light transport in the limit of very small imperfections, and by numerical results obtained for two waveguide geometries that offer contrasted damping behaviours.

The impact of random imperfections on the propagation of light in single–mode waveguides or fibers is critical for many applications. For weakly confined modes, the transport is dominated by absorption losses or by scattering into radiation modes. The incremental transmission loss per unit length is the same for any waveguide section, and the transmission $T$ decays exponentially with the waveguide length $L$, $\ln(T) = -\alpha L$, where $\alpha$ (often expressed in dB/cm) is referred to as the attenuation–rate coefficient [1].

Recently, the transport of slow light in periodic waveguides has retained much attention due to its potential for on–chip integration of time–domain optical processing [2]. Indeed, periodic waveguides such as photonic-crystal waveguides (PhCWs), represent a simple mean to finely control the speed of light by tuning the wavelength. Light transport of fast modes in periodic waveguides is similar to that in translation-invariant waveguides, but as the light velocity is slowed down, backscattering prevails, and one gradually moves from an "incoherent" loss damping to a coherent multiple scattering regime for which the phase of the wave field matters. The physical mechanisms in slow– and fast–light transports in monomode waveguides are thus totally different.

Despite this difference, since its very early studies at the beginning of the last decade [2-6], slow–light transport in periodic monomode waveguides has been assumed to follow the exponential damping law $\langle\ln(T)\rangle = -\beta L$ [7], and the attenuation-rate coefficient $\alpha$ is measured with the same classical methods as those used for translation-invariant waveguides, see [2,5] for example. The widely accepted belief that the attenuation $\langle\ln(T)\rangle$ of periodic waveguides linearly varies with $L$ prevails nowadays [8-11], and $\beta$ is presently considered as an important figure–of–merit that needs to be optimized. Even though $\beta$ was introduced in a confusing manner, more by analogy with translation-invariant waveguides than through a firm theoretical basis, the classical exponential damping law can be legitimated by early theoretical studies on the transport of electrons in one–dimensional (1D) metallic wires [12]. These studies have concluded that the attenuation is a self-averaging Gaussian–distributed quantity and that $\langle\ln(T)\rangle = -L/\ell$, with $\ell$ being the localization length [12-14] or more appropriately the damping length in the presence of losses.

As will be evidenced by computational results obtained for waveguides operating with moderately-low group-velocities and by analytic arguments derived in the perturbation limit, a strict exponential damping is not preserved in monomode periodic waveguides as one continuously tunes the group velocity of their fundamental modes. There are several reasons. First, the transport in periodic monomode waveguides is always a combination of several physical effects such as radiation in the cladding [6], interference between radiated and guided fields [8], so that in the end, the simplistic 1D models do not strictly apply. In addition, the exponential damping predicted in [12-14] is obtained in the scaling limit of $L \gg \ell$ and one may wonder if this limit is reached in practice with on-chip waveguides whose lengths rarely exceed a few millimeters. Even more importantly, we note that the exponential scaling law of localization is theoretically derived for specific assumptions on the scattering process, for instance that the fields scatter with a random phase uniformly distributed over $2\pi$ [12]. Nothing guaranties that such a randomization process, which seems realistic for translation invariant metallic wires and electrons, remains valid for photonic Bloch modes in periodic waveguides [15]; the geometry, the disorder model and the scattering physics, all matter [16].

We start by considering computational results of $<\ln(T)>$ for two different geometries, a photonic–crystal waveguide (PhCW) obtained by removing a single row of holes in a triangular lattice etched into a silicon ($n$ = 3.48) membrane, with a periodicity constant $a$ = 420 nm and a membrane thickness of 220 nm, and a sub-$\lambda$ grating nanowire (GNW) formed by a chain of silicon boxes embedded in a $SiO_2$ ($n$ = 1.44) host medium, see the insets in Fig. 1. Remarkably low propagation losses of 2.1 dB/cm in the telecom C–band have been recently reported [17] for this geometry. The segment dimensions are given in the caption, and the grating period is $a$ = 400 nm, corresponding to a $SiO_2$ 100–nm–gap length. For TE–like polarization (with $E_y$ = 0 on $y$ = 0), both waveguides support a truly–guided Bloch mode with a vanishing group velocity at the first Brillouin–zone boundary at telecommunication wavelengths, see Fig. 1.

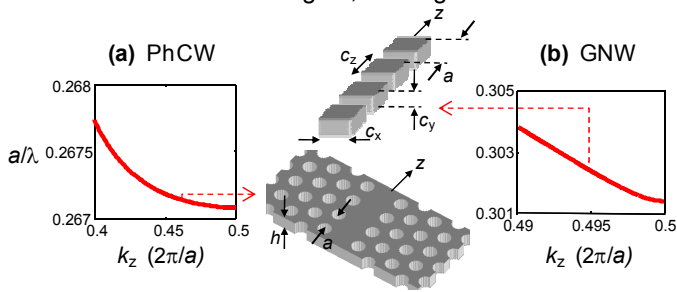

**FIG. 1 (color on line).** Dispersion relation of the waveguide geometries considered in the present work. The PhCW has a periodicity constant $a$ = 420 nm and a membrane thickness $h$ = 220 nm. The segment dimensions of the GNW are $c_x$ = 300 nm, $c_y$ = 260 nm and $c_z$ = 300 nm, and the period is $a$ = 400 nm, corresponding to a $SiO_2$ 100-nm-gap length. The band edges ($n_g \rightarrow \infty$) of the PhCW and GNW are located at $\lambda$ = 1.57 and 1.33 µm, respectively. Note that the $SiO_2$ host medium of the GNW is not depicted in the inset.

We assume a size disorder model for the PhCW, in which only the hole radii of the two inner rows (see Fig. 1) are randomly and independently varied around a mean value of $0.3a$ with a statistical Gaussian distribution. For the GNW, we assume that the wire etching results in unwanted Gaussian–distributed variations of the etched gaps, the period being constant. For both waveguides, we use the same standard deviation $\sigma$. The disorder models are likely to be simplistic but they represent a good compromise between computational complexity and fidelity to reality [18] and we believe that they do not impact our main conclusions. We further assume that every disorder realization of a waveguide of length $L = Na$ is defined by a random sequence of $N$ individual independent disordered single-cells, and that the fabrication process results in small deformations, which are totally independent from one cell to another (short–range disorder model). The waveguide transmission is calculated using a coupled–Bloch–mode method [8], which carefully considers the physics of the scattering process by taking into account out-of-plane leakage, in–plane multiple scattering at short distance between the perturbed holes of a single unit cell and at long distance between perturbed holes belonging to different cells.

We have calculated ensemble–averaged attenuations $<\ln(T^N)>$ for waveguide lengths $L = Na$, $N$ = 1, 2 … 10,000 and for several values of $n_g$, using the same realistic disorder levels $\sigma$ = 2 nm [11]. Figure 2a shows illustrative results obtained for $n_g$ = 30 by averaging over 1000 independent random waveguide realizations. In a log–scale, the PhCW (red–dashed curve) attenuation is linear in the length, right from $N$=5 to $N$=$10^3$, and even further (not shown) up to the maximum length $10^4 a$ considered in the computation. In contrast, one cannot define an attenuation rate coefficient for the GNW. The latter exhibits a nonlinear damping, with an initial rapid fall-off followed by an essentially linear variation starting only for $L > 500a$, a length roughly equal to 5 times the damping length. The energy–dependent damping rate observed at large $L$'s will be denoted by $\beta_\infty$ hereafter. This first numerical example invalidates the common belief that the attenuation coefficient of periodic waveguides is a constant that is independent of the waveguide length. We have repeated the calculations, systematically varying the group index and the disorder level. Figures 2b and 2c summarize the main useful results obtained for $\sigma$ = 2 nm and show the local damping rate $\beta_{loc}$, defined as $\beta_{loc} = -d<\ln(T^N)>/dL$. We have normalized $\beta_{loc}$ by $\beta_\infty$ to rescale our data and to better evidence deviation from an exponential damping. For the PhCW case shown in Fig. 2b, $\beta_{loc}$ is essentially equal to $\beta_\infty$ for all $L$ values ranging from $5a$ to $1000a$ and for all $n_g$'s ranging from 5 to 70, the small statistical fluctuations of $|\beta_{loc}/\beta_\infty|$ around 1 in the inset being due to the finite number of independent realizations used for averaging. This implies that $<\ln(T^N)>$ undergoes a linear variation in $L$ for all $n_g$'s, legitimating the definition of a frequency–dependent attenuation–rate coefficient $\beta$, which may even be used as a figure–of–merit to evaluate the performance of PhCWs, as has been done in previous works [2-6,8-11]. Although this can be somewhat expected from earlier works on localization in 1D systems [12-13], by no way the exponential damping can be considered as a triviality, especially for large $n_g$. In our opinion, it is even surprising. The GNW case in Fig. 2b offers radically different perspectives: $<\ln(T)>$ is no longer linear in $L$. The case is striking for large $n_g$'s; at $n_g$ = 70, $\beta_{loc}/\beta_\infty$ is as large as 4 for $L$ = 40$a$, which corresponds to a 3-dB attenuation on average, and keeps on decreasing over a few hundreds of periods, virtually preventing the observation of a stationary damping rate at typical scales encountered in integrated circuits. Note that similar results leading to similar conclusions have been obtained for other disorder levels, $\sigma$ = 1 or 3 nm, and for other PhCW disorder models with hole roughness instead of hole–size variation.

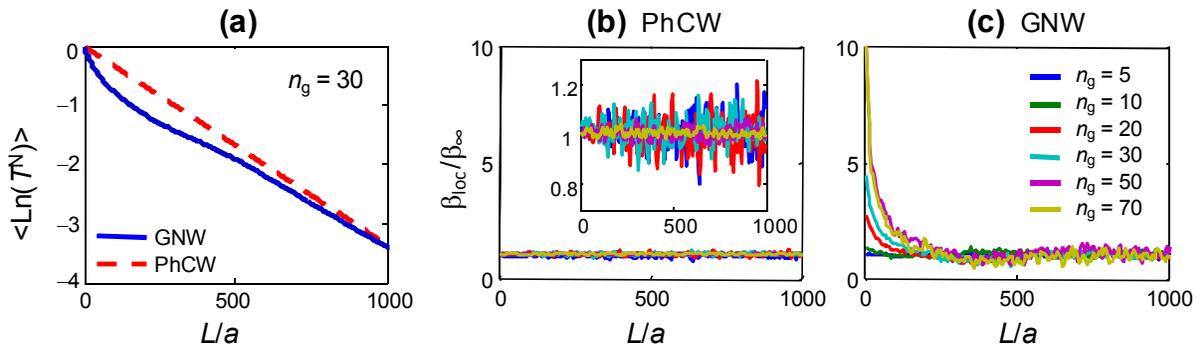

**FIG. 2 (color on line).** Testing the validity of the exponential damping law. **(a)** $<\ln(T^N)>$ as a function of the waveguide length $L$ for $n_g = 30$. **(b)** and **(c)** The normalized local attenuation–rate $\beta_{loc}/\beta_\infty$ from $L = 5a$ (≈ 2 μm) to $1000a$ (≈ 500 μm) for several $n_g$, see the legend in **(c)**. **(b)** PhCW. **(c)** GNW. The inset in **(b)** is an enlarged view essentially showing that the exponential damping is satisfied for PhCWs and that $\beta \approx \beta_\infty$ for all $n_g$. Note that $\beta_\infty$ indeed depends on $n_g$ and $\sigma$, $\beta_\infty \propto (\sigma n_g)^2$ [6,8]. All plots are obtained for $\sigma = 2$ nm. The rapid oscillations in **(b)** and **(c)** are artifacts resulting from an ensemble-averaging over a finite number of $10^3$ independent realizations.

To get an insight, the plan is to derive a recursive relationship between the ensemble–averaged attenuation of waveguides of length $L$ and $L+a$. This is possible only if the multiple–scattering of *radiated* waves, which could be first excited by a scatterer and then recycled back into the waveguide by another nearby scatterer, is neglected. With this approximation, the energy transport is solely due to the fundamental Bloch mode of the periodic waveguide and a 2×2 scattering–matrix formalism can be used. Knowing the transmission and rear–reflection coefficients $t^N$ and $r^N$ of a disordered waveguide of length $Na$ and the elementary scattering coefficients $\tau_{N+1}$ and $\rho_{N+1}$ of the unit cell $N+1$, see Fig. 3, the law of composition of scattering matrices [19] leads to the recursive formula $t^{N+1} = t^N \tau_{N+1} / (1-\rho_{N+1} r^N)$. Taking the logarithm and averaging, one straightforwardly obtains the master relationship for $T^N = |t^N|^2$

$$<\ln(T^{N+1})> - <\ln(T^N)> = <\ln|\tau_{N+1}|^2> - <\ln|1-\rho_{N+1} r^N|^2>. \quad (1)$$

Equation (1) straightforwardly supports our main finding, i.e. the initial nonlinear behavior followed at large $L$'s by an essentially linear variation with a constant damping rate $\beta_\infty$. For short–range and uniform disorders, all unit cells belong to the same population and the probability distributions of $\tau_{N+1}$ and $\rho_{N+1}$ are independent of $N$. In contrast, the coefficient $r^N$ in the rightmost term does depend on $N$. Thus the damping rate is not constant as it depends on $L$. However, as $L$ increases, a stationary regime appears: adding new unit cells just lowers the transmission but no longer impacts the reflected light ($r^N$ saturates to a value $r^\infty$ independent of $N$, as shown later on), and the damping rate becomes constant, $\beta_\infty = -a^{-1}<\ln|\tau_{N+1}/[1-\rho_{N+1} r^\infty]|^2>$.

The remaining of the Letter is devoted to explaining why the waveguides have different attenuation behavior, why the PhCW exhibits a fully linear behavior and under which condition one may observe such a behavior that allows for a straight measurement and definition of attenuation rates. To this end, it is advisable to start by considering the asymptotic case in which the disorder is very small. In the perturbation regime, because $|r^N| < 1$ and $|\rho_{N+1}| << 1$ for all $N$'s, $|\rho_{N+1} r^N| << 1$, and with a first–order Taylor expansion, the $L$–dependent term on the right side of Eq. (1) becomes $-<\ln|1-\rho_{N+1} r^N|^2> \approx <\rho_{N+1} r^N + (\rho_{N+1} r^N)^*>$. The rear–reflection coefficient $r^N$ of the waveguide of length $L = Na$ depends on all the elementary scattering coefficients $\rho_n$, $n = 1, 2 \ldots N$. As mentioned before, for short–range and uniform disorders, all the $\rho_n$'s have the same probability distribution and are independent from each other, so

$$<\ln(T^{N+1})> - <\ln(T^N)> \approx <\ln|\tau_{N+1}|^2> + <\rho><r^N> + <\rho^*><(r^N)^*>, \quad (2)$$

where the average value of $\rho_{N+1}$ is denoted $<\rho>$ since it is independent of $N$. Consistently, $<\ln|\tau_{N+1}|^2>$ in Eq. (1) is denoted by $<\ln|\tau|^2>$. Still in the perturbation regime, $\rho$ is known analytically, $\rho = \int d\mathbf{r}\, \alpha(\mathbf{r})\Delta l(\mathbf{r})$ [20], where the integral runs over the perturbed boundaries $\mathbf{r}$ of the unit cell, $\alpha(\mathbf{r})$ is a complex function that depends on the Bloch–mode electric–field distribution and that takes into account local field corrections, and $\Delta l(\mathbf{r})$ is the deformation. Since $<\Delta l(\mathbf{r})> = 0$ for any $\mathbf{r}$ by definition, we straightforwardly obtain that $<\rho> = 0$, and Eq. (1) becomes

$$<\ln(T)> = <\ln|\tau|^2>L, \quad (3)$$

for any $L$. Actually, we conclude that waveguide attenuations are linear in their lengths in the perturbation regime. Indeed this result that holds whatever the length (i.e. smaller, equal or larger than the damping length) provides a theoretical support for our common belief. We believe that, except if an intentional disorder is provided, classical $z$–invariant waveguides, such as high-index-contrast nanowires, operate in the perturbation regime and obey the classical linear damping law. From Figs. 2b and 2c, it is reasonable to expect similar conclusions for fast light transport in periodic waveguides manufactured with state of the art fabrication processes.

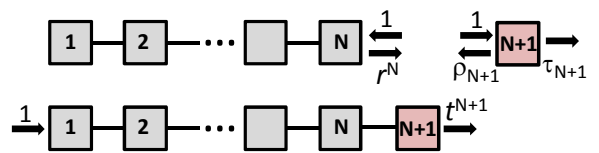

**FIG. 3 (color on line).** Definition of the main scattering coefficients involved in the 1D model. $t^N$ and $r^N$ denote the transmission and rear-reflection coefficients of a disordered waveguide of length $Na$ and $\tau_{N+1}$ and $\rho_{N+1}$ are the elementary scattering coefficients of the unit cell $N+1$. For short–range and uniform disorders, the $\rho_m$'s (or $\tau_m$'s) all belong to the same population and are all independent.

In view of the above, the nonlinear behavior of the GNW at large $n_g$'s testifies the breakdown of the perturbative regime, manifesting itself in a deviation of $\langle\rho\rangle$ from zero. However, the marked contrast between the GNW and PhCW responses is not simply due to a difference in $\langle\rho\rangle$ since the two waveguides have comparable $\rho$'s: As $n_g$ increases from 10 to 30, and then to 70, $|\langle\rho\rangle| = 0.001$, 0.017 and 0.097 for the PhCW and 0.002, 0.017 and 0.055 for the GNW. These values all remain much smaller than 1, legitimating the use of the approximate Eq. (2) out of the perturbation regime. We now argue that the exponential damping law breaks down for the GNW because $|\langle r^N\rangle|$ is much larger for the GNW than for the PhCW, see the blue and red curves in Fig. 4. Although multiple scattering significantly impacts the complex oscillatory behavior of $|\langle r^N\rangle|$, the essence of the effect can be simply captured by neglecting the denominator term in the classical formula, $r^{N+1} = \rho_{N+1} + r^N \tau_{N+1}^2/(1+\rho_{N+1} r^N)$ [19]. In this way one readily obtains the series $\langle r^{N+1}\rangle \approx \langle\rho\rangle + \langle r^N\rangle \langle\tau^2\rangle$, showing that $\langle r^N\rangle = 0$ in the perturbation regime. Beyond this regime, the solution $\langle r^N\rangle$ of the recursive equation follows a spiral–like path in the complex plane, which explains the oscillatory behavior in Fig. 4. It tends to $\langle r^\infty\rangle = \langle\rho\rangle/(1-\langle\tau^2\rangle)$, whose magnitude can be much larger than $\langle\rho\rangle$ for $\langle\tau^2\rangle \approx 1$. Since the GNW slow modes lie much closer to the first Brillouin–zone boundary, where $k_z = \pi/a$ and hence $\langle\tau^2\rangle \approx \exp(2ik_z a) = 1$, than those of the PhCW, $\langle r^\infty\rangle$ is expected to be larger for the GNW than for the PhCW, consistently with Fig. 4. Physically, this implies that, in the vicinity of the Brillouin–zone boundary, because the elementary backscattered waves add essentially in phase (intuitively, this still holds if multiple scattering is considered), $\langle r^N\rangle$ takes large values and $\langle \ln(T^N)\rangle$ exhibits a spatial transient regime until the mean and the higher moments of $r^N$ converge. Finally, let us stress that a small value of $|\langle r^N\rangle|$ does not imply that the backreflected intensity $\langle|r^N|^2\rangle$ is small. Actually, the attenuation of the PhCW is smaller than that of the GNW for small $n_g$'s, is equal for $n_g = 30$ (Fig. 2a), and is larger at lower group velocities.

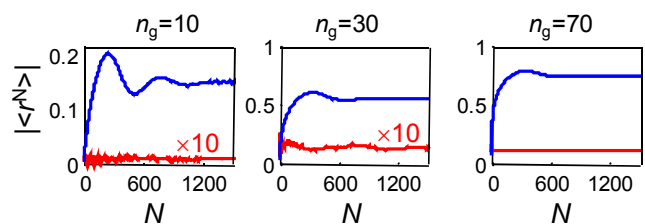

**FIG. 4 (color on line).** Modulus of the averaged rear–reflection coefficients $r^N$ as a function of $L = Na$ for the PhCW (red), the GNW (blue), and for $n_g = 10$, 30 and 70. Note that $r^N$ is expected to be null on ensemble–average in the limit of small perturbations. The results hold for $\sigma = 2$ nm and are obtained by averaging over $10^4$ independent realizations of disorder.

In summary, the transport of *fast* light in monomode periodic waveguides manufactured with state–of–the–art fabrication tools is likely to obey the classical exponential damping law. This classical property does not remain valid as one continuously lowers the speed of light in the waveguide. Before reaching a stationary damping–rate regime as $N\to\infty$, a transient regime with much larger local damping–rates is first observed in general. The transient length strongly depends on the scattering processes and on the waveguide geometry, and may reach a few hundreds of periods even for moderately–small slowdown factors of 20 ($n_g \approx 70$). We therefore conclude that the estimation of the attenuation–rate or even the localization length of periodic waveguides should be handled with caution, especially if intentional disorder is introduced to study strong perturbation cases.

The authors thank J.P. Hugonin and Kevin Vynck for fruitful discussions. SM thanks the DGA for his PhD fellowship. WS thanks the Triangle de la Physique. The work is supported by the ANR project CALIN.

———————